\documentclass[3p,times,procedia]{elsarticle}

\usepackage{ecrc}

\volume{00}

\firstpage{1}

\journalname{Physics Procedia}

\runauth{}

\jid{phpro}

\jnltitlelogo{Physics Procedia}

\CopyrightLine{2011}{Published by Elsevier Ltd.}

\usepackage{amssymb}
\usepackage[figuresright]{rotating}

\newcommand{\epm}{\ensuremath{e^{\pm}\;}}

\newcommand{\omb}{\ensuremath{\Omega_{\rm B} h^{2}}}
\newcommand{\neff}{\ensuremath{{\rm N}_{\rm eff}}}
\newcommand{\Deln}{\ensuremath{\Delta{\rm N}_\nu}}

\newcommand{\mchi}{\ensuremath{m_\chi}}
\def\lsim{\lesssim}  
\def\gsim{\gtrsim}

\newcommand{\beq}{\begin{equation}}  
\newcommand{\eeq}{\end{equation}}

\def\3he{$^3$He}
\def\4he{$^4$He}
\def\7li{$^7$Li}
\def\Yp{Y$_{\rm P}$}

\def\hii{H\thinspace{$\scriptstyle{\rm II}$}}

\newcommand{\eg}{{\it e.g.}}

\begin{document}

\begin{frontmatter}



\dochead{}
%% e.g. \dochead{17th International Conference on Dynamical Processes in Excited States of Solids}

\title{Light WIMPs And Equivalent Neutrinos}


\author[gs]{Gary Steigman} 
\ead{steigman.1@osu.edu}

\address[gs]{Physics Department, Center for Cosmology and Astro-Particle Physics, The Ohio State University, Columbus, Ohio, USA\\
Departamento de Astronomia, Universidade de S\~ao Paulo,
S\~ao Paulo, Brasil}

\author[kn]{Kenneth M. Nollett}
\ead{nollett@ohio.edu}

\address[kn]{Department of Physics and Astronomy, Ohio University, Athens, Ohio, USA}

\begin{abstract}
Very light WIMPs ($\chi$), thermal relics that annihilate late in the early Universe, change the energy and entropy densities at BBN and at  recombination.  BBN, in combination with the CMB, can remove some of the degeneracies among light WIMPs and equivalent neutrinos, constraining the existence and properties of each.  Depending on the nature of the light WIMP (Majorana or Dirac fermion, real or complex scalar) the joint BBN + CMB analyses set {\bf lower} bounds to \mchi~in the range $0.5 - 5\,{\rm MeV}$ ($m_{\chi}/m_{e} \gsim 1 - 10$), and they identify {\bf best fit} values for \mchi~in the range $5 - 10\,{\rm MeV}$.  The {\bf joint} BBN + CMB analysis finds a {\bf best fit} value for the number of equivalent neutrinos, \Deln~$\approx 0.65$, nearly independent of the nature of the WIMP.   In the absence of a light WIMP (\mchi~$\gsim 20\,{\rm MeV}$), \neff~$= 3.05(1 + \Deln/3)$.  In this case, there is excellent agreement between BBN and the CMB, but the joint fit reveals \Deln~$= 0.40\pm0.17$, disfavoring standard big bang nucleosynthesis (SBBN) (\Deln~= 0) at $\sim 2.4\,\sigma$, as well as a sterile neutrino (\Deln~= 1) at $\sim 3.5\,\sigma$.   The best BBN +  CMB joint fit disfavors the absence of dark radiation (\Deln~= 0 at $\sim 95\%$ confidence), while allowing for the presence of a sterile neutrino (\Deln~= 1 at $\lsim 1\,\sigma$).  For all cases considered here, the lithium problem persists.  These results, presented at the TAUP 2013 Conference, are based on \citet{kngs}.

\end{abstract}


\end{frontmatter}

% \linenumbers

\section{Introduction And Overview}
\label{intro}

In the absence of ``extra", equivalent neutrinos (dark radiation) or light ($\lsim 20\,{\rm MeV}$), weakly interacting massive particles (WIMPs), the particle content relatively late in the early Universe is quite simple.  After the \epm pairs (and all the other more massive standard model (SM) particles) have annihilated ($T \lsim m_{e}$), the only remaining SM particles are the CMB photons and the three SM neutrinos ($\nu_{e}$, $\nu_{\mu}$, $\nu_{\tau}$).  At these early epochs the Universe is ``radiation dominated" and the energy density may be written as $\rho_{\rm R} = \rho_{\gamma} + 3\,\rho_{\nu}$, where $3\,\rho_{\nu}$ accounts for the contributions from the three SM neutrinos.  During these early epochs the contributions to the total mass/energy density from the baryons (B) and the dark matter (DM), as well as any dark energy (DE), are very subdominant compared to $\rho_{\rm R}$.  More generally, in addition to the SM neutrinos, there may be extra, ``beyond the standard model" particles that, like the SM neutrinos, are extremely light ($\lsim 10\,{\rm eV}$) and very weakly interacting.  During the early (or, even, relatively late) evolution of the Universe these neutrino-like particles, so called ``equivalent neutrinos", will contribute to the energy density.  The energy density controls the early Universe expansion rate.  If \Deln~counts the contribution of equivalent neutrinos, often referred to as ``dark radiation", $\rho_{\rm R} = \rho_{\gamma} + (3 + \Deln)\,\rho_{\nu}$.  The contribution to \Deln~of an equivalent neutrino that decouples along with the SM neutrinos (at $T = T_{\nu d}$) will be \Deln~= 1 for a Majorana fermion (\eg, a sterile neutrino), \Deln~= 2 for a Dirac fermion or, \Deln~= 4/7 for a real scalar.  In general, \Deln~is an integer (fermions) or an integer multiple of 4/7 (bosons).  However, an equivalent neutrino that is more weakly interacting than the SM neutrinos, will have decoupled earlier in the evolution of the Universe and its contribution to \Deln~will be suppressed by the heating of the SM neutrinos (and photons) when the heavier SM particles decay and/or annihilate.  Therefore, in principle, there is no reason that \Deln~should be an integer or an integer multiple of 4/7 (for further discussion see \citet{steig13}; for a specific example of three, very weakly coupled, right-handed neutrinos, see \citet{haim} and for the example of a weakly coupled scalar particle see \citet{weinberg}).

After the SM neutrinos decouple, when $T = T_{\nu d} \approx 2 - 3\,{\rm MeV}$, the \epm pairs annihilate, heating the photons but not the neutrinos.  Prior to neutrino decoupling (and \epm annihilation), the neutrinos, \epm pairs, and the photons are in equilibrium at the same temperature, $T_{\nu} = T_{e} = T_{\gamma}$ but, after \epm annihilation, the photons are hotter than the relic neutrinos (or, equivalently, the neutrinos are cooler than the photons).  In most simplified analyses it is assumed that the neutrinos decoupled instantaneously and that the electrons were effectively massless at neutrino decoupling.  With these approximations the late time (after \epm annihilation is complete) ratio of neutrino and photon temperatures, $(T_{\nu}/T_{\gamma})_{0} = (4/11)^{1/3}$, follows from entropy conservation.  The late time ratio of the energy density in one species of neutrino ($\rho^{0}_{\nu}$) to that in the photons is $(\rho^{0}_{\nu}/\rho_{\gamma})_{0} = 7/8\,(T_{\nu}/T_{\gamma})^{4}_{0} = 7/8\,(4/11)^{4/3}$.  However, at neutrino decoupling $m_{e}/T_{\nu d} \approx 0.2 \neq 0$ and, as a result, $\rho_{\nu}$ differs (by a small amount) from $\rho^{0}_{\nu}$ \citep{steig13}.  Furthermore, the neutrinos don't really  decouple instantaneously.  While the neutrinos are partially coupled to the annihilating \epm pairs they share a small amount of the energy released by the annihilation \citep{dolgov,dolgov-osc,enqvist,hannestad,mangano}.  These effects can be accounted for by introducing \neff, the ``effective number of neutrinos" where, at late times ($T_{0} \ll m_{e}$), $\rho_{\rm R\,0} \equiv \rho_{\gamma\,0} + {\rm N}_{\rm eff}\,\rho^{0}_{\nu\,0}\,,$
\beq
{\rm N}_{\rm eff} = 3\,\bigg[{11 \over 4}\bigg({T_{\nu} \over T_{\gamma}}\bigg)^{3}_{0}\bigg]^{4/3}\bigg(1 + {\Deln \over3}\bigg)\,.
\eeq
It should be kept in mind that while the relative contributions of neutrinos and photons to the total radiation density may be evolving before and during BBN, \neff~is a ``late time" quantity, evaluated long after BBN has ended, when the only relativistic particles remaining are the photons and the neutrinos.

Under the assumptions of instantaneous neutrino decoupling and $m_{e} \ll T_{\nu d}$, \neff~= 3 + \Deln.  Keeping the instantaneous decoupling approximation but correcting for the finite electron mass, \neff~$\approx 3.02(1 + \Deln/3)$ \citep{steig13}.  Following the non-instantaneous neutrino decoupling and allowing for the finite electron mass, \neff~$\approx 3.05(1 + \Deln/3)$ \citep{mangano}.  It should be noted that in this latter case there is, in addition, a very small, but not entirely negligible correction to the BBN predicted primordial helium abundance \citep{mangano}.  Since the expansion rate (Hubble parameter) of the radiation dominated early Universe is $H \propto \rho_{\rm R}^{1/2}$, the presence of dark radiation (\Deln~$\geq 0$) results in a speed up to the early Universe expansion rate.  

\begin{figure}[!t]
\begin{center}
\includegraphics[width=0.45\columnwidth]{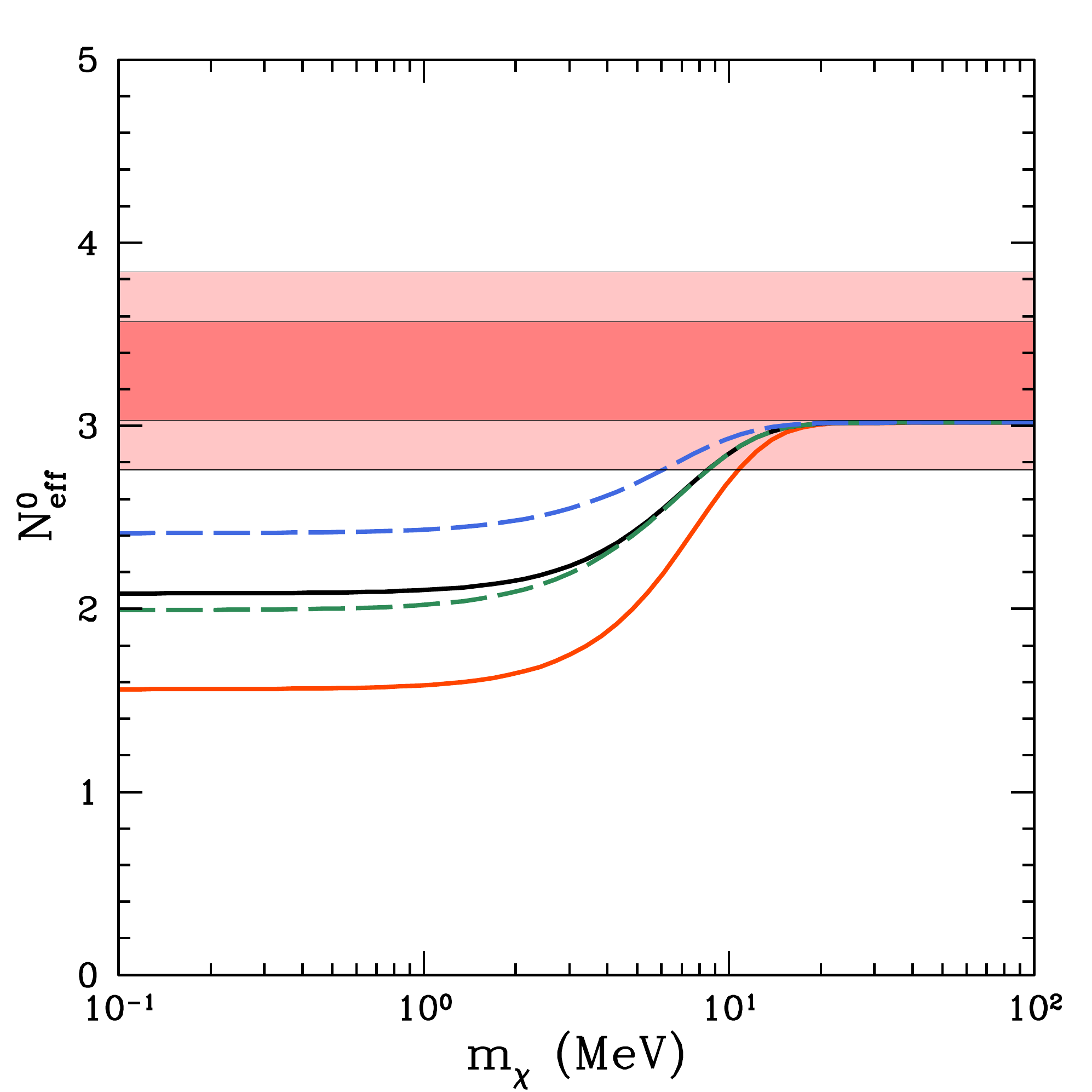}
\hskip .4in
\includegraphics[width=0.45\columnwidth]{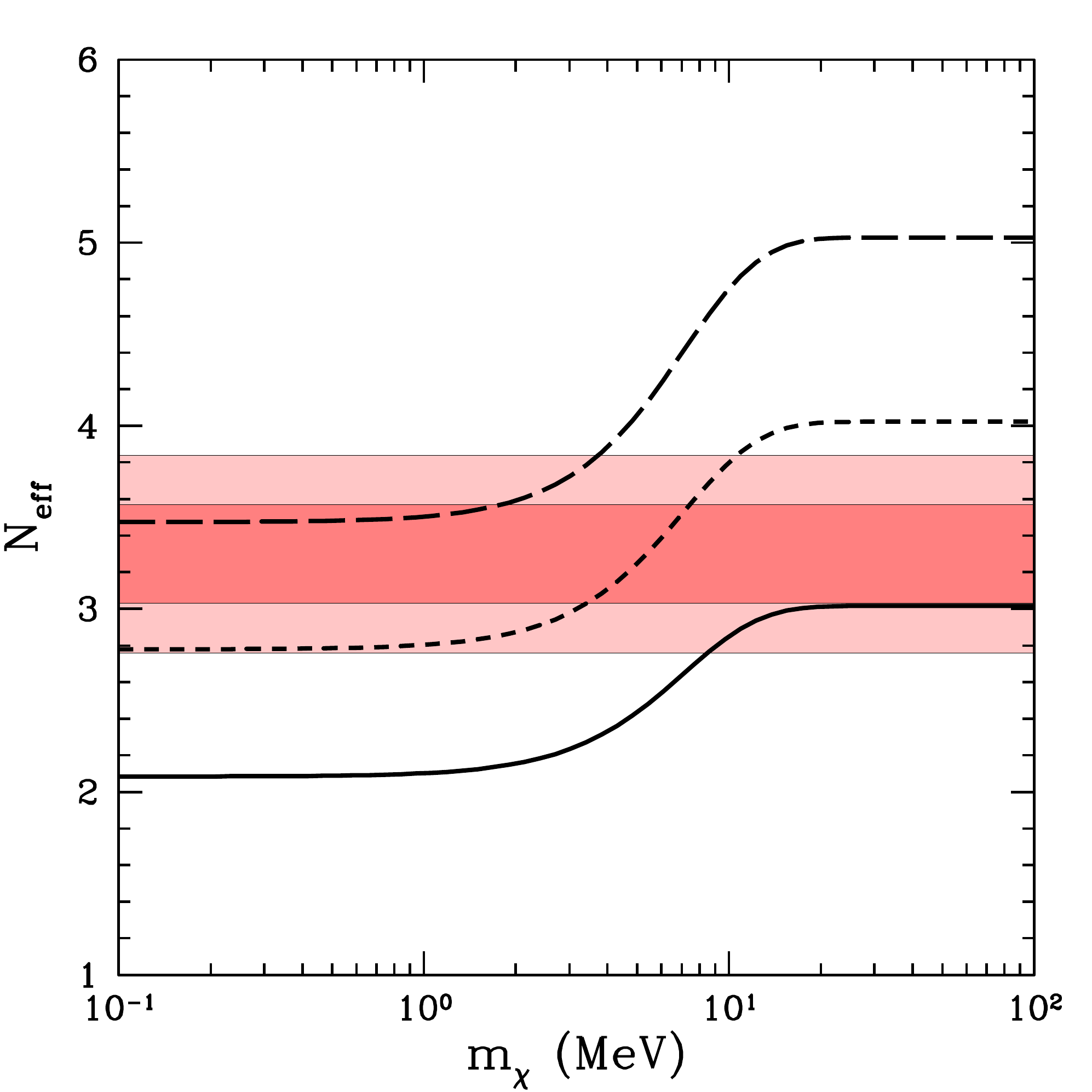}
%\\\vskip 0.2in
\caption{The left panel shows N$^{0}_{\rm eff}$ (\neff~when \Deln~= 0) as a function of the WIMP mass for electromagnetically coupled light WIMPs in the absence of equivalent neutrinos.  From bottom to top, the solid red curve is for a Dirac WIMP,  the dashed green curve is for a complex scalar, the solid black curve is for a Majorana fermion and, the dashed blue curve is for a real scalar.  The horizontal, red/pink bands are the Planck CMB 68\% and 95\% allowed ranges for N$_{\rm eff}$.  The right panel specializes to the case of a Majorana fermion WIMP, showing N$_{\rm eff}$ as a function of the WIMP mass for \Deln~equivalent neutrinos.  The solid  curve is for $\Delta {\rm N}_{\nu} = 0$, the short dashed curve is for $\Delta {\rm N}_{\nu} = 1$ and, the long dashed curve is for $\Delta {\rm N}_{\nu} = 2$.  The horizontal red bands are the Planck CMB 68\% and 95\% allowed ranges for N$_{\rm eff}$, including baryon acoustic oscillations in the CMB constraint.}
\label{fig:neffvsmchi}
\end{center}
\end{figure}

So far, the possibility of a very light, weakly interacting, massive particle, a WIMP $\chi$, has been ignored in the discussion here.  The difference between a WIMP and an equivalent neutrino is that a WIMP remains thermally coupled to the SM particles until after it has become non-relativistic and begins annihilating.  As a result, the light WIMP annihilation heats the remaining SM particles (either the photons and, possibly, the \epm pairs if the WIMP couples electromagnetically or, the SM neutrinos if the WIMP only couples to them).  Note that in the analysis and discussion here, the WIMP {\bf need not} be a dark matter candidate; the WIMP could be a sub-dominant component of the dark matter ($\Omega_{\chi} < \Omega_{\rm CDM}$).  For example, the WIMP could be a light, millicharged particle such as that proposed by \citep{millicharged} and discussed by Dolgov at this conference.  Here, as in \citet{kngs}, we specialize to the case of a light WIMP coupled only to the photons and \epm pairs.  The relevant role for BBN and the CMB played by such a light WIMP is that its annihilation heats the photons relative to the decoupled SM neutrinos, changing (reducing) $(T_{\nu}/T_{\gamma})_{0}$.  After \epm and WIMP annihilation, at fixed photon temperature, the neutrinos, SM and equivalent, are cooler than the photons.  In this case, \neff~is a function of \mchi~(see \citet{steig13} and references therein).  Since the expansion rate of the early Universe is controlled by the energy density, any modification of \neff~will be reflected in a non-standard expansion rate (\eg, during BBN and at recombination).  In addition, extremely light WIMPs (\mchi~$\lsim m_{e}$) will annihilate so late that, if their annihilation produces photons, they will modify the baryon-to-photon ratio ($\eta_{10} = 10^{10}(n_{\rm B}/n_{\gamma})_{0} = 273.9\,\omb$) during or after BBN.  BBN can probe \neff~(through the effects of the neutrinos on the expansion rate and on the weak rates that regulate the neutron to proton ratio) as well as the universal ratio of baryons-to-photons.  At late times in the early Universe, \eg, at recombination, the CMB can also probe \omb~and \neff.  As independent probes of the effective number of neutrinos (\neff) or the number of equivalent neutrinos (\Deln) and the universal baryon density (\omb~or $\eta_{10}$), BBN and the CMB can help to break the degeneracies among these parameters and the WIMP mass (and spin/statistics) and to constrain their allowed ranges (see, \citet{steig13} \& \citet{kngs} and Fig.\,\ref{fig:neffvsmchi}).

\subsection{Planck CMB Constraints}

In their analysis, \citet{kngs}, whose results are described and summarized here, adopted the CMB constraints on \omb~and \neff~from the Planck $\Lambda\mathrm{CDM}+N_\mathrm{eff}$ fit including supplementary baryon acoustic oscillation (BAO) data \citep{planck}.  The correlations between these quantities were included in \citep{kngs} and in the analysis here.  From the CMB we adopted \omb~$= 0.0223 \pm 0.0003$ ($\eta_{10} = 6.11 \pm 0.08$) and \neff~$= 3.30 \pm 0.27$.  In Fig.\,\ref{fig:neffvsmchi}, the Planck 68\% and 95\% constraints on \neff~are shown as a function of the WIMP mass (the CMB constraints are independent of the WIMP mass).  Also shown are the curves corresponding to \neff~as a function of \mchi~for a Majorana fermion WIMP and for three choices of the number of equivalent neutrinos.  The behavior seen here is qualitatively similar for a Dirac or scalar WIMP (see, \eg, \citep{steig13} \& \citep{kngs}).  This figure illustrates the degeneracies between \neff~and \mchi.  For example, for \Deln~= 0 the CMB can set a {\bf lower} bound to \mchi.  In contrast, for \Deln~= 1 or 2, it is high values of \mchi~that are excluded.

\subsection{BBN Constraints}

\begin{figure}[!t]
\begin{center}
\includegraphics[width=0.55\columnwidth]{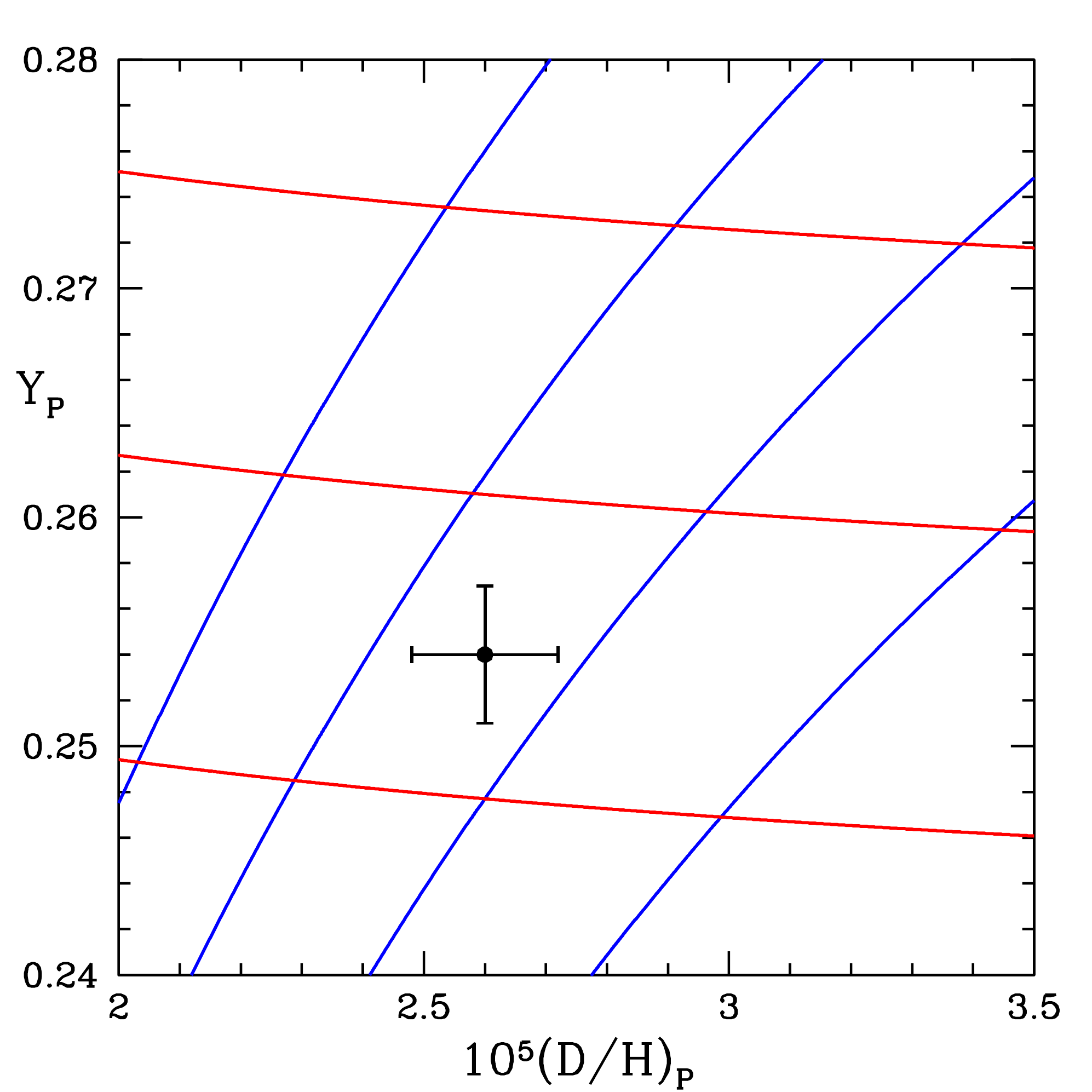}
\caption{BBN predicted curves of constant baryon-to-photon ratio and equivalent number of neutrinos in the \Yp~-- $y_{\rm DP}$ plane.  From left to right (blue), $\eta_{10} = 7.0, 6.5, 6.0, 5.5$.  From bottom to top (red) \Deln~= 0, 1, 2.  Also shown by the filled circle and error bars are the observationally inferred values of \Yp~and $y_{\rm DP}$ adopted here (see the text).}
\label{fig:YvsD}
\end{center}
\end{figure}

Of the light nuclides produced during BBN, D and \4he are the relic nuclei of choice.  To account for, or minimize, the post-BBN modifications of the primordial abundances, observations at high redshift (z) and/or low metallicity (Z) are preferred.  Deuterium (and hydrogen) is observed in high-z, low-Z, QSO absorption line systems and helium is observed in relatively low-Z, extragalactic \hii~regions.  Even so, it may still be necessary to correct for any post-BBN nucleosynthesis that may have modified their primordial abundances.  The post-BBN evolution of D and \4he is simple and monotonic.  As gas is cycled through stars, D is destroyed and \4he produced.  Finally, D and \4he provide complementary probes of the parameters (\Deln~and \omb) of interest.  $y_{\rm DP} \equiv 10^{5}{\rm (D/H)}_{\rm P}$ is mainly sensitive to the baryon density at BBN (\omb) and is less sensitive to \Deln.  In contrast, the \4he mass fraction, \Yp, is very insensitive to \omb, but is quite sensitive to \Deln.  This complementary, nearly orthogonal, dependence of D and \Yp~on $\eta_{10}$ and \Deln~is illustrated in Fig.\,\ref{fig:YvsD}.  For the analysis here (and in \citet{kngs}), we have adopted, $y_{\rm DP} = 2.60 \pm 0.12$ \citep{pettini} and \Yp~$= 0.254 \pm 0.003$ \citep{izotov}. 

In contrast, the post-BBN evolution of \3he and \7li, the other two nuclides produced in significant abundances during BBN, is complex and model dependent.  \3he has only been observed in the relatively metal-rich interstellar medium of the Galaxy and its BBN-predicted abundance is less sensitive to \omb~and \Deln~than is the BBN D abundance.  \3he is not used in our BBN analysis, but we have confirmed that its observationally inferred primordial abundance \citep{rood} is in good agreement with our BBN-predicted results.  \7li suffers from some of the same issues as \3he.  Although \7li is observed in very metal poor stars, its post-BBN evolution is complicated and model dependent, especially the connection between the surface lithium abundances observed and those in the gas out of which the stars formed.  While in principle \7li could be as useful as D in constraining \omb~(and, to a lesser extent, \Deln), there is the well known ``lithium problem" (see, \eg, \citet{fields} and \citet{spite} for recent reviews) that, as will be seen below, persists.  In the BBN analyses, with and without a light WIMP, only D and \4he are used to constrain \omb~and \Deln~(or, \neff) and these BBN constraints are compared to the independent constraints from the CMB.

\section{BBN With A Light WIMP}
\label{sec:bbn+LW}

\begin{figure}[!t]
\begin{center}
\includegraphics[width=0.55\columnwidth]{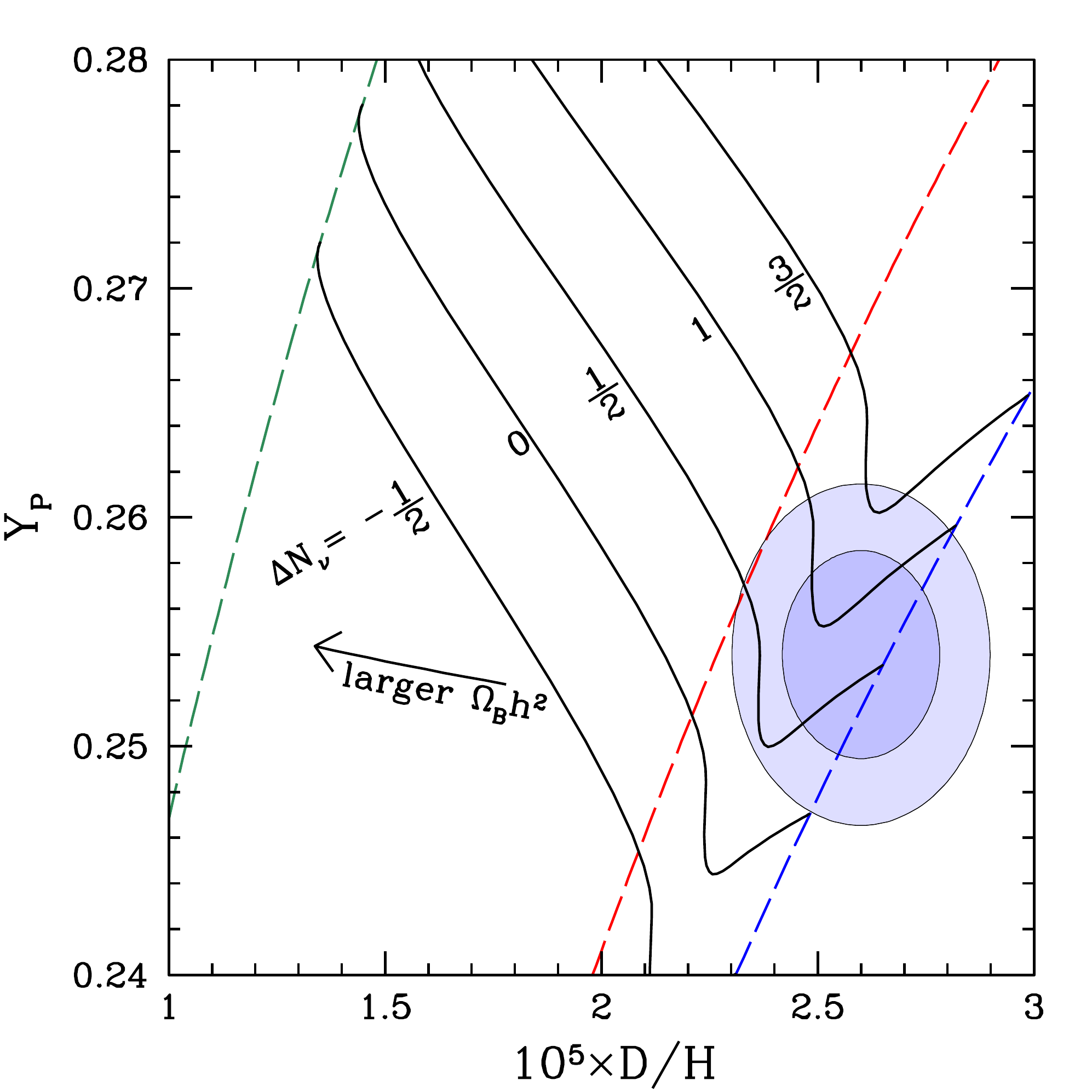}
  \caption{BBN yields of D/H and \Yp, computed for Majorana WIMPs at \omb~$= 0.022$ for several values of \Deln~($-1/2 \leq \Deln \leq 3/2$) as labeled.   Along each curve of fixed \Deln, $m_\chi$ varies from $\infty$ ($m_{\chi} \gg 20\,{\rm MeV}$) at the right end to 10 keV at the left end.  The dashed curves show the yields at fixed $m_\chi$ but varying \Deln~with $m_\chi\rightarrow\infty$ (blue, right side), $m_\chi = m_e$ (red, middle), and $m_\chi = 0$ (green, left side).  The arrow shows the effect of increasing the baryon density.  The observational constraints Y$_{\rm P} = 0.254\pm 0.003$ and $y_\mathrm{D} = 2.60\pm 0.12$ are shown as 68\% (darker) and 95\% (lighter) joint confidence contours.}
 \label{fig:wimpyields-majorana}
 \end{center}
\end{figure}

As will be seen below in \S\,\ref{sec:bbn+nLW}, BBN and the CMB are in excellent agreement in the absence of a light WIMP.  Here, the main interest is in investigating the constraints BBN and the CMB can set on the mass of such a WIMP, and how its presence changes the constraints on \neff~(\Deln) and \omb.  The presence of a light WIMP can effect BBN (and the CMB) in several ways, provided it is sufficiently light.  For example, a very light WIMP might be mildly relativistic at BBN (or, prior to BBN, when the neutron-to-proton ratio is being set), contributing to the total energy density (similar to an equivalent neutrino) and speeding up the expansion rate.  A faster expansion generally increases the neutron-to-proton ratio at BBN, leading to the production of more \4he.  Also, such a very light WIMP might annihilate during or after BBN and the photons produced by its annihilation will change the baryon-to-photon ratio from its value during BBN.  The baryon-to-photon ratio at present may differ from its value at BBN affecting, mainly, the BBN D abundance.  The effects on the BBN light element yields in the presence of a light WIMP, neglecting any equivalent neutrinos (\Deln~$\equiv 0$), were investigated by \citet{ktw} and \citet{serpico} and, more recently, by \citet{boehm}.  In \citet{kngs} those BBN calculations were extended to allow for the presence of dark radiation (\Deln~$\neq 0$).  In this case, there are three free parameters.  In addition to the baryon density ($\eta_{10}$ or \omb) and the number of equivalent neutrinos (\Deln), the light WIMP mass is allowed to vary, modifying the connection between \Deln~and the late time quantity \neff,
\beq
\neff~= {\rm N}_{\rm eff}^{0}(\mchi)(1 + \Deln/3)\,,
\eeq
and producing time-dependent effects on the weak rates and the expansion rate during BBN.  As already noted by \citet{ktw}, \citet{serpico}, and \citet{boehm}, for an electromagnetically coupled light WIMP, as \mchi~decreases below $\sim$\,$20\,{\rm MeV}$, the BBN predicted D abundance decreases monotonically, while the \4he abundance first decreases (very slightly) and then increases monotonically.  This behavior is shown in Fig.\,\ref{fig:wimpyields-majorana}.  For a more detailed discussion of the physics controlling this modified BBN, especially the non-monotonic behavior of \Yp~and its connection to the temperature dependence of the neutron -- proton interconversion reactions, see \citet{kngs}.

\begin{figure}[!t]
\begin{center}
\includegraphics[width=0.45\columnwidth]{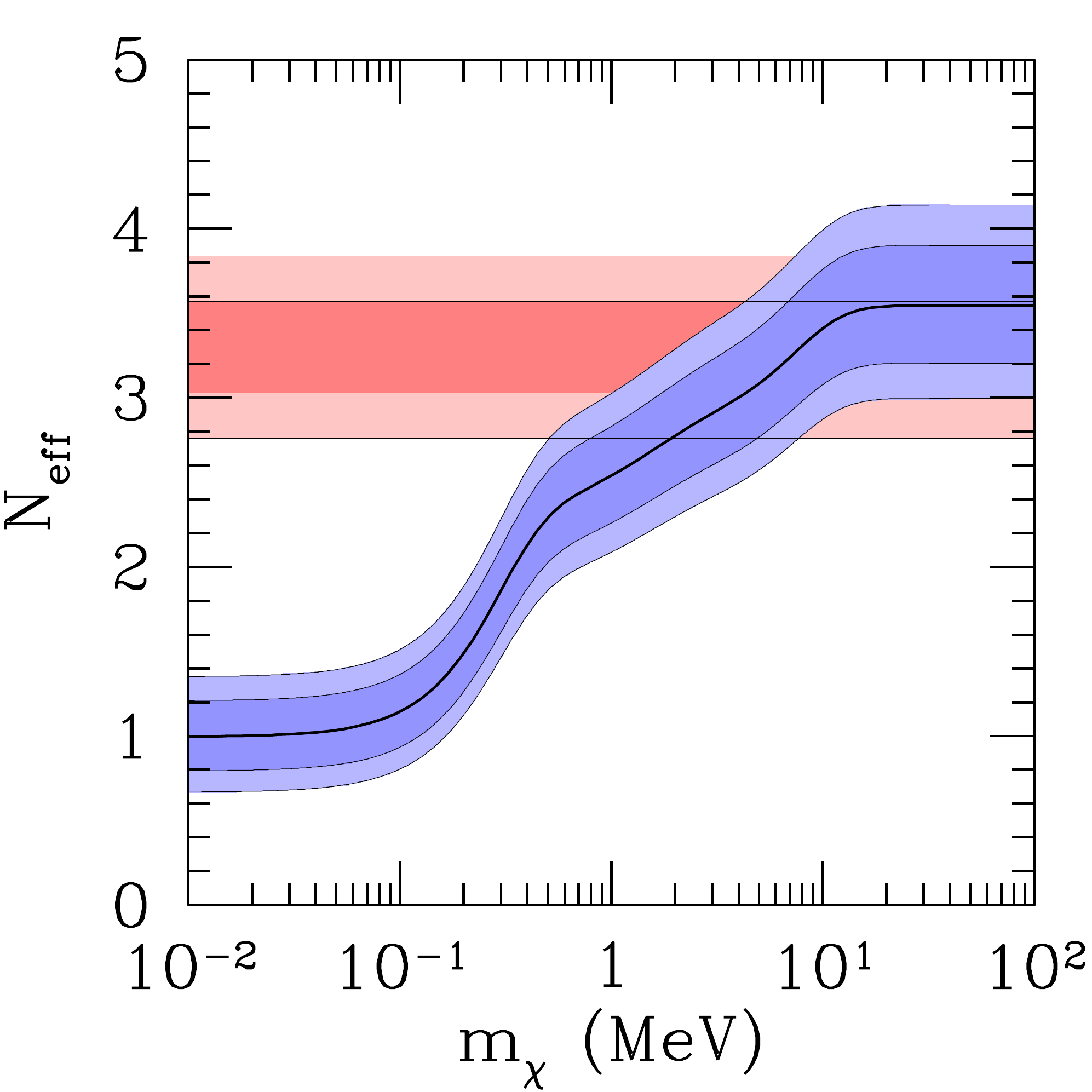}
\hskip 0.4 in
\includegraphics[width=0.45\columnwidth]{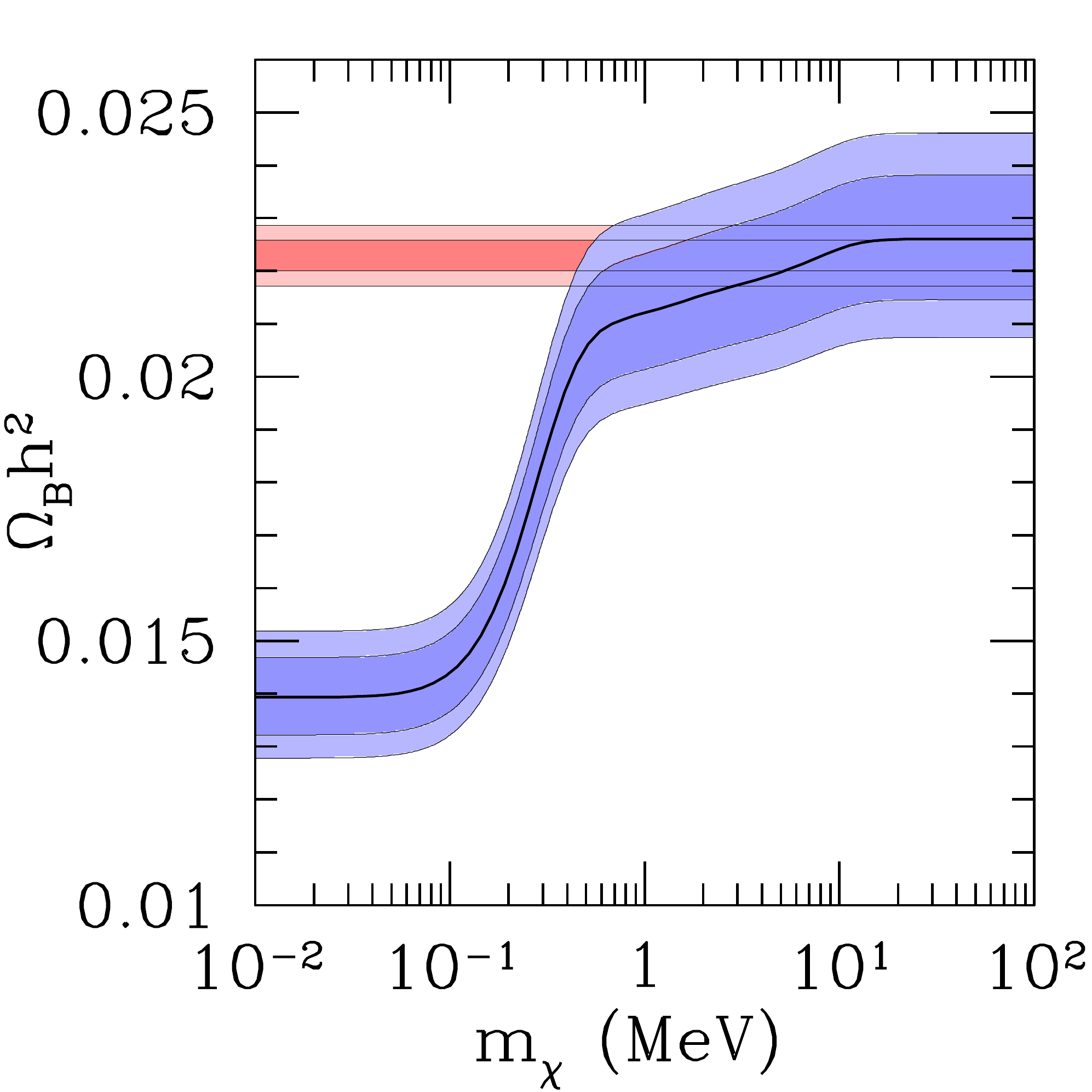}
\caption{In the left hand panel are shown the CMB and BBN constraints on \neff~as a function of the light WIMP mass, \mchi.  The horizontal, pink bands show the 68\% and 95\% ranges from the Planck CMB results.  The blue bands show the corresponding BBN ranges.  The black curve through the middle of the blue bands shows the values of \neff~as a function of \mchi~for which the BBN predicted D and \4he abundances agree exactly with the observationally inferred abundances adopted here.  The right hand panel shows the corresponding results for the baryon mass density, \omb, as a function of the WIMP mass.}
\label{fig:neffvsmchi1}
\end{center}
\end{figure}

With three parameters and only two observables ($y_{\rm DP}$ and \Yp), BBN is underconstrained.  For each choice of \mchi, a pair of $\eta_{10}$ (\omb) and \Deln~(\neff) parameters can be found so that BBN predicts -- exactly -- the observed D and \4he abundances.  This is illustrated in the two panels of Fig.\,\ref{fig:neffvsmchi1}, which show \neff~(left panel) and \omb~(right panel) as functions of the WIMP mass, as inferred from the CMB (where \neff~and \omb~are independent of \mchi) and from BBN.  These figures show how the degeneracy illustrated in Fig.\,\ref{fig:neffvsmchi} can be broken by combining constraints from the CMB with those from BBN.  

\begin{figure}[!t]
\begin{center}
\includegraphics[width=0.45\columnwidth]{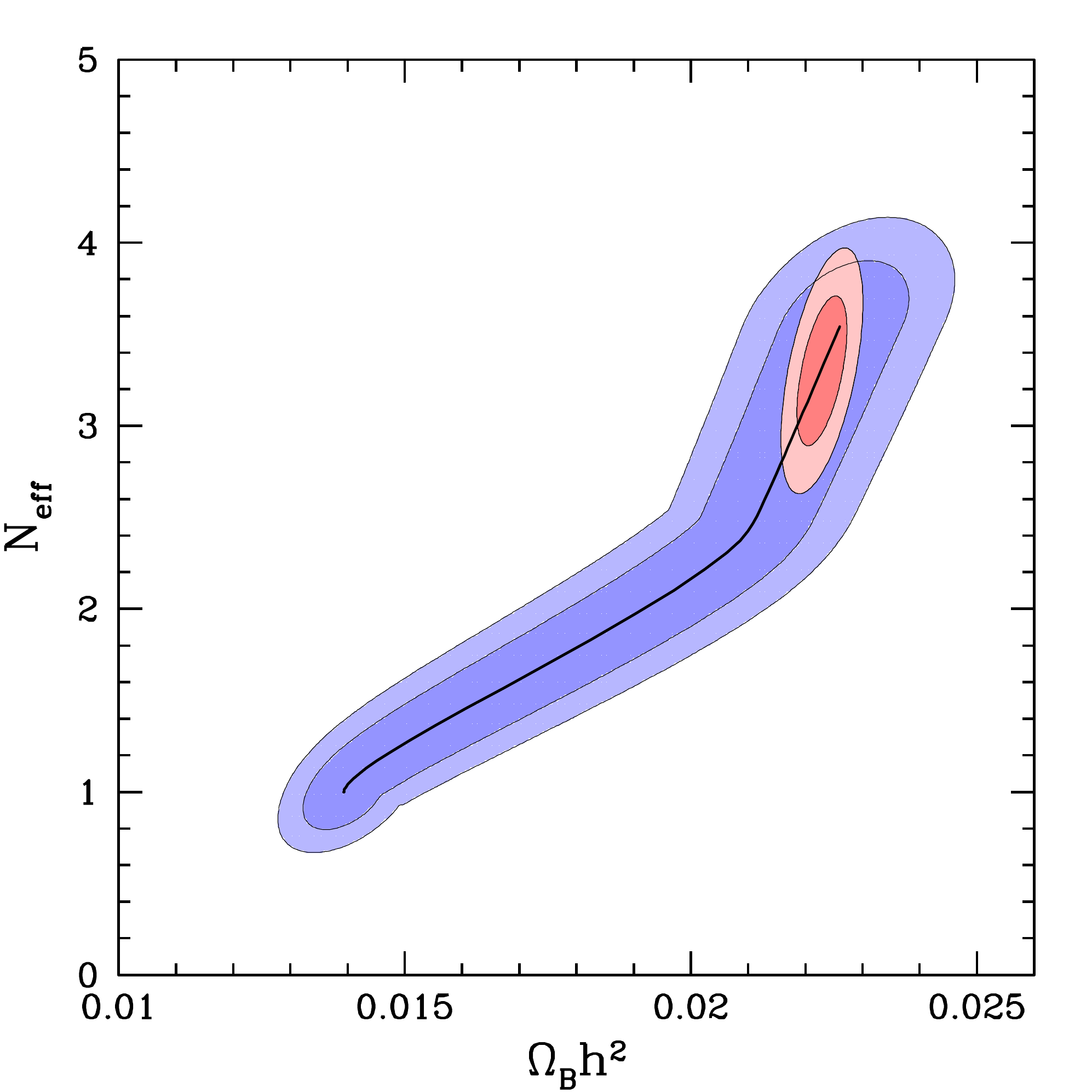}
\hskip 0.4 in
\includegraphics[width=0.45\columnwidth]{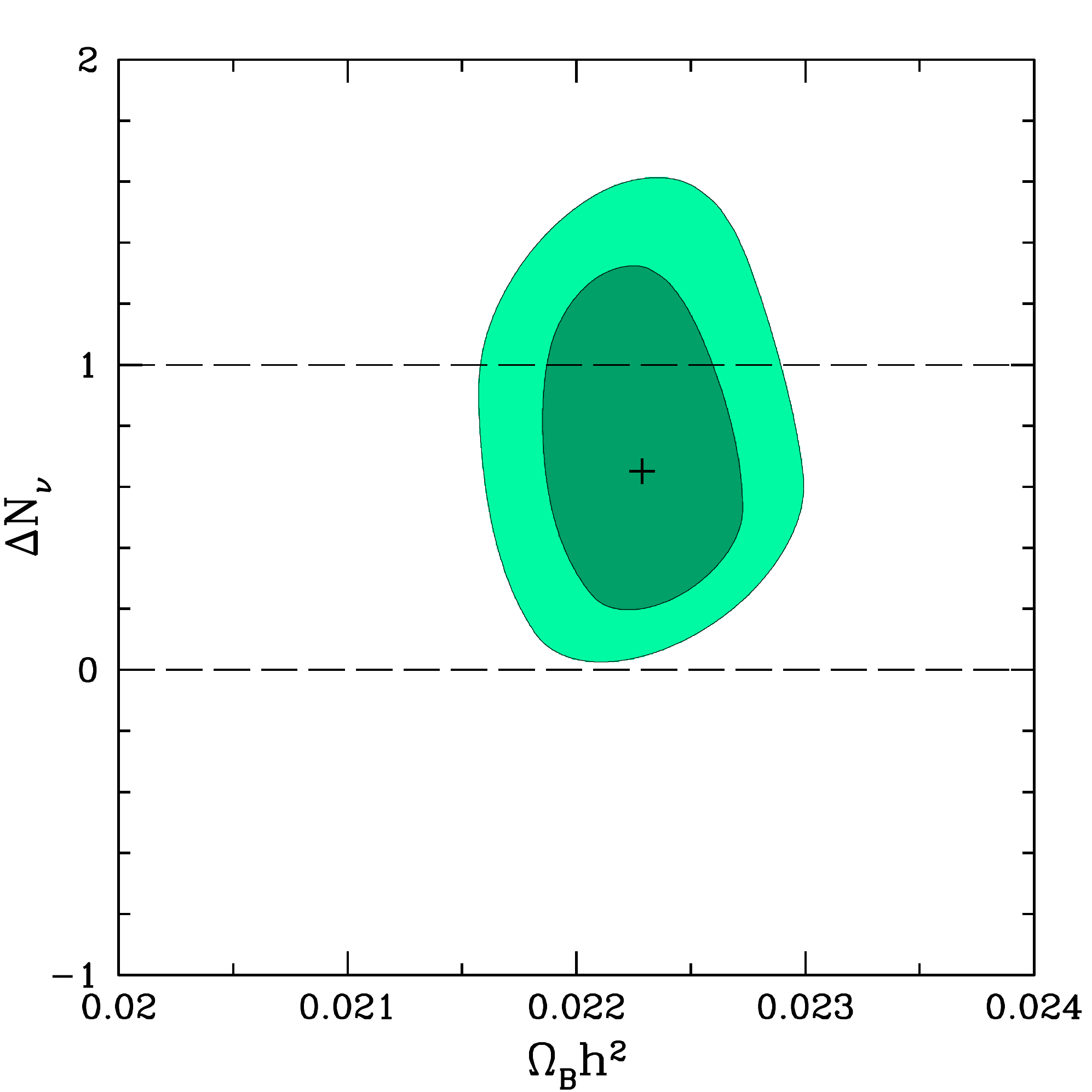}
\caption{The left hand panel shows the CMB and BBN constraints on \neff~as a function of the the baryon density \omb~(combining the results shown in the two panels of Fig.\,\ref{fig:neffvsmchi1}).  The right hand panel shows the combined BBN + CMB constraint on dark radiation (\Deln) as a function of the baryon mass density (\omb) in the presence of a light WIMP.}
\label{fig:neffvsomb}
\end{center}
\end{figure}

A comparison of the BBN and CMB constraints on \neff~and \omb~is shown in Fig.\,\ref{fig:neffvsomb}.  Over the range in \neff~and \omb~defined by the Planck CMB constraints, the independent and complementary BBN and CMB results are in excellent agreement.  As a result, the BBN and CMB results may be combined in a joint analysis to identify the 68\% and 95\% ranges allowed in the \neff~(or, \Deln) -- \omb~(or, $\eta_{10}$) plane.  This joint analysis \citep{kngs} finds \neff~$= 3.30 \pm 0.26$ and \omb~$= 0.0223 \pm 0.0003$ ($\eta_{10} = 6.11 \pm 0.08$), consistent with the CMB results alone, but with slightly smaller uncertainties.  The new results from this joint analysis for \Deln~as a function of \omb~are shown in the right hand panel of Fig.\,\ref{fig:neffvsomb}.  For the joint fit, \Deln~$= 0.65^{+0.46}_{-0.35}$.  Note that these figures and the numerical results cited here are for the case of a Majorana fermion WIMP.  Very similar results are found for a Dirac fermion or for a real or complex scalar WIMP (see Table 1 in \citet{kngs}).

Allowing for a light WIMP (Majorana or Dirac fermion, real or complex scalar), the joint CMB + BBN analysis excludes light WIMPs with masses $\lsim 0.5 - 5\,{\rm MeV}$.  The best joint fit WIMP mass is found to be in the range \mchi~$\approx 5 - 10\,{\rm MeV}$, depending on the nature of the WIMP.  However, very nearly independently of the nature of the WIMP, the best fit for the dark radiation is \Deln~$\approx 0.65$ (see Fig.\,10, Table 1, and the related discussion in \citet{kngs}).  In all cases, the best fit for the effective number of neutrinos is \neff~$= 3.30$ (N$_{\rm eff}^{0} \approx 2.71$).  While \Deln~= 0 is still disfavored at $\sim$\,$95\%$ confidence, in the presence of a light WIMP, a sterile neutrino (but, not two sterile neutrinos!) is now permitted.  Since the no light WIMP case is a good fit to the BBN and CMB data (see \S\,\ref{sec:bbn+nLW}), there is no upper bound to the WIMP mass.

It is noteworthy that for the WIMP masses allowed by the joint BBN + CMB fit (including the high WIMP mass limit -- the no light WIMP case), the BBN predicted lithium abundance lies in the range ${\rm A(Li)} \equiv 12 + {\rm log\,(Li/H)} = 2.72 \pm 0.04$ (see Fig.\,13 in \citet{kngs}), still a factor of $\sim 3$ larger than the observationally inferred Spite Plateau value of A(Li) $= 2.20 \pm 0.06$ \citep{spite}.  A light WIMP does not help to alleviate (indeed, it reinforces) the lithium problem.  It is not surprising that the lithium problem persists since the D and \7li abundances, set at the same time during BBN, are (anti)correlated and the primordial deuterium abundance is used as to constrain the parameters required by BBN to predict the lithium abundance.

\section{BBN Without A Light WIMP}
\label{sec:bbn+nLW}

\begin{figure}[!t]
\begin{center}
\includegraphics[width=0.45\columnwidth]{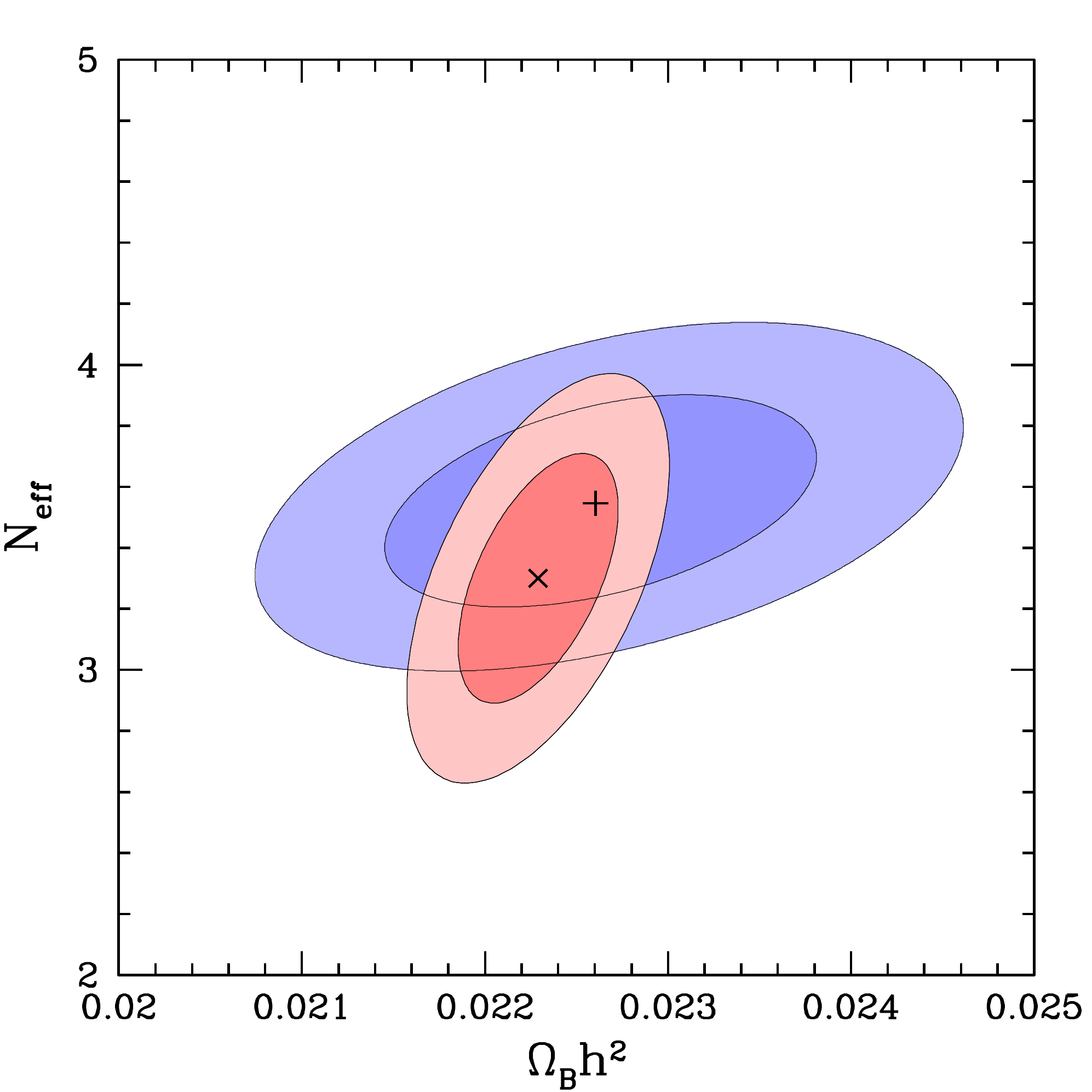}
\hskip 0.4 in
\includegraphics[width=0.45\columnwidth]{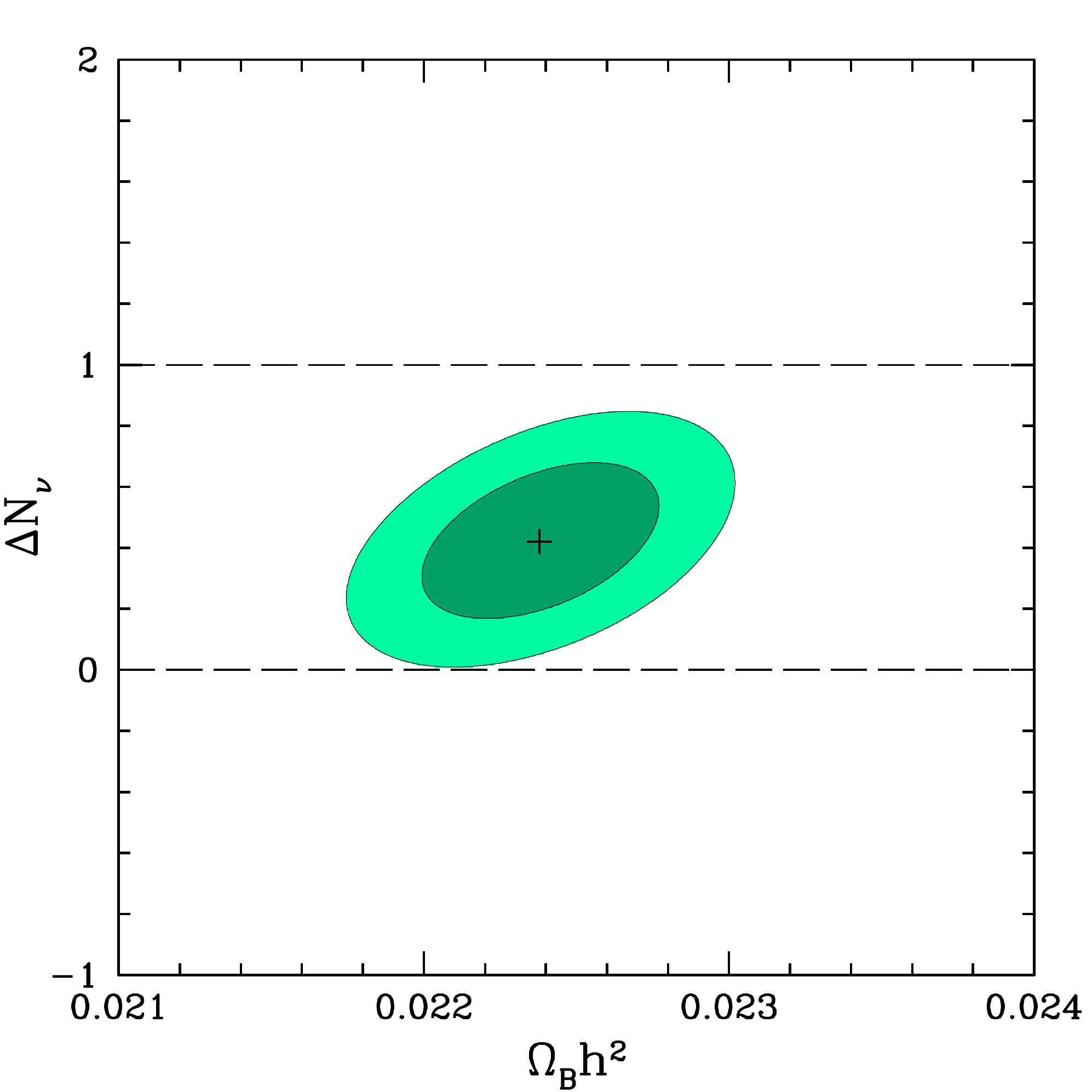}
\caption{The left hand panel shows a comparison of the the 68\% (darker) and 95\% (lighter) contours in the \neff~--~\omb~plane derived separately from BBN (blue) and the CMB (pink).  The ``$\times$'' symbol marks the best fit CMB point and the ``$+$'' is the best fit BBN point.  In the right hand panel the combined BBN + CMB constraint on dark radiation (\Deln) is shown as a function of the baryon mass density (\omb).}
\label{fig:neffvsomb0}
\end{center}
\end{figure}

In the absence of a light WIMP (or other non-standard physics) the BBN-predicted primordial abundances depend on only two parameters, the baryon-to-photon ratio ($\eta_{10}$ or, the baryon mass density \omb) and the number of equivalent neutrinos (\Deln).  In the absence of a light WIMP the effective number of neutrinos and the number of equivalent neutrinos are related by \neff~$ = 3.05\,(1 + \Deln/3)$.  With two, independent, relic abundances (D and \4he), BBN can constrain these two parameters as illustrated in Fig.\,\ref{fig:YvsD}.  For the abundances adopted here, BBN (without a light WIMP) predicts, $\eta_{10} = 6.19 \pm 0.21$ (\omb~$= 0.0226 \pm 0.0008$) and \Deln~$= 0.51 \pm 0.23$, corresponding to \neff~$= 3.56 \pm 0.23$ (accounting for round-off).  The BBN 68\% and 95\% contours in the \neff~-- \omb~plane, along with the best fit point, are shown in Fig.\,\ref{fig:neffvsomb0}, where they are compared to the corresponding contours (and best fit point) for these parameters inferred from the independent Planck CMB data \citep{planck}.  As the left hand panel of Fig.\,\ref{fig:neffvsomb0} reveals, in the absence of a light WIMP, there is excellent agreement between BBN and the CMB.  This motivates (justifies) a joint BBN + CMB analysis, resulting in (for the joint fit) $\eta_{10} = 6.13 \pm 0.07$ (\omb~$= 0.0224 \pm 0.0003$) and \neff~$= 3.46 \pm 0.17$ (\Deln~$= 0.40 \pm 0.17$).  However, as may be seen from the right hand panel of Fig.\,\ref{fig:neffvsomb0}, this joint BBN + CMB fit favors neither standard BBN (SBBN: \Deln~= 0), nor the presence of a sterile neutrino (\Deln~= 1).  SBBN is disfavored at $\sim 2.4\,\sigma$ and a sterile neutrino is disfavored at $\sim 3.5\,\sigma$.  

As for lithium, even without a light WIMP, for the joint BBN + CMB parameter values, the BBN predicted \7li abundance is ${\rm A(Li)} = 2.72 \pm 0.03$.  The lithium problem, the factor of $\sim$\,3 difference between predictions and observations, persists.

\section{Summary And Conclusions}

In the absence of a light WIMP and equivalent neutrinos (no dark radiation), BBN (SBBN) depends on only one parameter, the baryon abundance.  For the adopted primordial D and \4he abundances, SBBN predicts a best fit baryon density of $\eta_{10} = 6.0 \pm 0.2$ \citep{kngs}, in excellent agreement with the corresponding value of the baryon abundance ($\eta_{10} = 6.06 \pm 0.07$) inferred from the Planck analysis with \neff~fixed \citep{planck}.  However, as shown in \citet{kngs}, for either of these baryon abundances, BBN predicts \Yp~$= 0.247\,(\pm\, 0.0005)$, which is a poor fit to the observationally inferred primordial helium abundance of \Yp~$= 0.254\pm0.003$ \citep{izotov}.  

In the absence of a light WIMP, but now allowing for dark radiation (\Deln~and \neff~free to vary), the effective number of neutrinos and the number of equivalent neutrinos are related by \neff~$= 3.05(1 + \Deln/3)$.  From the Planck CMB analysis alone, \neff~$= 3.30 \pm 0.27$ \citep{planck}, constraining the number of equivalent neutrinos to  \Deln~$= 0.25 \pm 0.27$, consistent with the absence of dark radiation (\Deln~= 0) at $\lsim 1\,\sigma$ and, inconsistent with a sterile neutrino (\Deln~= 1) at $\sim$\,$2.8\,\sigma$.  In addition to the constraint on \Deln, the CMB alone also provides a constraint on the universal baryon density, \omb~$= 0.0223 \pm 0.0003$ ($\eta_{10} = 6.11 \pm 0.08$).  For the Planck determined combination of the baryon abundance and the number of equivalent neutrinos, the BBN predicted D and \4he abundances ($y_{\rm DP} = 2.56 \pm 0.08$ and \Yp~$= 0.2505 \pm 0.0005$) are in very good agreement with the observationally inferred primordial abundances, while lithium (A(Li)~$= 2.73 \pm 0.03$) is too high.  For the same case (no light WIMP, \Deln~free to vary), the BBN fit to the observed D and \4he abundances is in excellent agreement with the CMB inferred parameter values (BBN: $\eta_{10} = 6.19 \pm 0.21$, \Deln~$= 0.51 \pm 0.23$, \neff~$= 3.56 \pm 0.23$).  A joint BBN + CMB analysis predicts $\eta_{10} = 6.13 \pm 0.07$ (\omb~$= 0.0224 \pm 0.0003$) and \Deln~$= 0.40 \pm 0.17$ (\neff~$= 3.46 \pm 0.17$).  BBN with these joint fit parameter values predicts $y_{\rm DP} = 2.60 \pm 0.08$ and \Yp~$= 0.2525 \pm 0.0005$.  The corresponding lithium abundance, A(Li)~$= 2.72 \pm 0.03$, is a factor of $\sim$ 3 higher than the observationally inferred, primordial value.

In the absence of a light WIMP, BBN and the CMB are in excellent agreement, but neither \Deln~$= 0$ (SBBN) nor \Deln~$= 1$ (a sterile neutrino) is favored by BBN or by the combined BBN + CMB fit.  In the presence of a sufficiently light WIMP (\mchi~$\lsim 20\,{\rm MeV}$) the CMB results are unchanged, although the connection between \neff~and \Deln~is modified depending on the WIMP mass, \neff~$= {\rm N}^{0}_{\rm eff}(\mchi)(1 + \Deln/3)$.  Now there is a degeneracy between the CMB constraints on \neff~and \Deln~(and \mchi).  As may be seen from Fig.\,\ref{fig:neffvsmchi}, for some choices of \Deln, the CMB constraint on \neff~sets a lower limit to \mchi, while for other choices the CMB sets an upper limit to the WIMP mass.  The independent constraints from BBN help to break these degeneracies.  In the presence of a light WIMP BBN now depends on three parameters: \Deln, \neff, \omb~(or, \Deln, \mchi, $\eta_{10}$), but there are only two BBN constraints (from D and \4he).  For each choice of \mchi, corresponding to a fixed value of ${\rm N}_{\rm eff}^{0}$, there is always a pair of \Deln~and $\eta_{10}$ values for which BBN predicts -- exactly -- the observationally inferred primordial D and \4he abundances adopted here.  However, the corresponding BBN inferred values (and ranges) of \neff~and \omb~need not necessarily agree with the values (and ranges) set by the CMB.  By comparing the BBN and CMB constraints, the degeneracies may be broken, leading to a lower bound, as well as a best fit value, of the WIMP mass (depending on the nature of the WIMP).  For the case of a Majorana fermion WIMP shown in the figures here, \mchi~$\gsim 1.7\,{\rm MeV}$ and the best fit is for \mchi~$ = 7.9\,{\rm MeV}$.  Depending on the nature of the WIMP, the lower bound to \mchi~ranges from $\sim$\,$m_{e}$ to $\sim$\,$10\,m_{e}$, while the best fit WIMP masses lie in the range $\sim$\,$5 - 10\,{\rm MeV}$ (see \citet{kngs}).  In all cases, very nearly independent of the nature of the WIMP, N$_{\rm eff}^{0} \approx 2.71$ and \Deln~$\approx 0.65$.  While the joint BBN + CMB analysis is dominated by the CMB values for \neff~and \omb, the presence of an additional free parameter (\mchi) relaxes the constraints (increases the error) on \Deln~compared to the no light WIMP case.  In the presence of a sufficiently light WIMP a sterile neutrino is now permitted at $\lsim 68\%$ confidence (see the right hand panel of Fig.\,\ref{fig:neffvsomb}).  However, the absence of dark radiation (\Deln~= 0) remains disfavored at $\sim 95\%$ confidence.  For the joint BBN + CMB analysis the BBN predicted primordial lithium abundance is A(Li)~$= 2.73 \pm 0.03$, essentially identical to that for the no light WIMP case.  The persistence of the lithium problem is largely a result of the strong coupling between the BBN predicted abundances of D and \7li, and cannot be resolved by an extension of SBBN to include equivalent neutrinos (\Deln~$\neq 0$) or light WIMPs.

It should be noted that since TAUP 2013, \citet{cooke} published new results on the primordial abundance of deuterium, $y_{\rm DP} = 2.53 \pm 0.04$.  Although their new central value agrees very well with the earlier, \citet{pettini} result adopted here, the new uncertainty is smaller by a factor of three.  In the analysis described here (and, in more detail in \citet{kngs}), this small change in the primordial deuterium abundance has the effect of increasing $\eta_{10}$ by $\sim$\,$0.1$ and decreasing \Deln~by $\sim$\,$0.01$.  These small changes, well within the current errors, leave the results and conclusions presented here unaffected.



\begin{center}
{\bf Acknowledgements}\\
\end{center}

We are grateful to the Ohio State University Center for Cosmology and Astro-Particle Physics for hosting K. M. N's visit during which most of the work described here was done.  K. M. N is pleased to acknowledge support from the Institute for Nuclear and Particle Physics at Ohio University.  G. S. is grateful for the hospitality provided by the Departamento de Astronomia of the Instituto Astron$\hat{\rm o}$mico e Geof\' \i sico of the Universidade de S\~ao~Paulo, where these proceedings were written.  The research of G. S. was supported at OSU by the U.S.~DOE grant DE-FG02-91ER40690.  

%% References


\bibliographystyle{elsarticle-num}
\bibliography{<your-bib-database>}


\bibliographystyle{aa}




% \end{thebibliography}

\end{document}